\documentclass[a4paper,11pt]{article}
\pdfoutput=1 
\usepackage{jheppub} 

\def\CN{{\cal N}}

\newcommand{\ket}[1]{ | {#1} \rangle }

\def\diag{\mathop{\rm diag}\nolimits}

\def\SO{\mathop{\rm SO}}

\def\SU{\mathop{\rm SU}}
\def\U{\mathrm{U}}
\def\Sp{\mathop{\rm Sp}}

\def\tr{\mathop{\rm tr}}

\def\su{\mathfrak{su}}

\def\sp{\mathfrak{sp}}

\def\beq#1\eeq{\begin{align}#1\end{align}}

\def\node#1{\overset{#1}{\circ}}
\def\blacknode#1{\overset{#1}{\bullet}}

\def\blackver{\overset{\displaystyle\bullet}{\scriptstyle\vert}}

\title{Instanton operators and symmetry enhancement in \\ 5d supersymmetric quiver gauge theories}

\preprint{}

\author[]{Kazuya Yonekura}
\author[]{}
\affiliation[]{School of Natural Sciences, Institute for Advanced Study, 1 Einstein Drive, \\
Princeton, NJ 08540, USA}

\abstract{
We consider general 5d $\SU(N)$ quiver gauge theories whose nodes form an ADE Dynkin diagram of type $G$.
Each node has $\SU(N_i)$ gauge group of general rank, Chern-Simons level $\kappa_i$ and additional $w_i$ fundamentals.
When the total flavor number at each node is less than or equal to $2N_i-2|\kappa_i|$,
we give general rules under which the symmetries associated to instanton currents are enhanced to $G \times G$ or a subgroup of it
in the UV 5d superconformal theory. When the total flavor number violates that condition at some of the nodes, further enhancement
of flavor symmetries occurs. In particular we find a large class of gauge theories interpreted as $S^1$ compactification of 6d superconformal theories
which are waiting for string/F-theory realization. We also consider hypermultiplets in (anti-)symmetric representation.
}

\begin{document} 
\maketitle
\flushbottom

\section{Introduction and summary}
By now, overwhelming evidence has been accumulated 
that there exist nontrivial superconformal field theories (SCFTs) in five and six dimensions, despite the lack of 
renormalizable interacting Lagrangians (other than $\phi^3$) in these dimensions.
In five dimensions, it was pioneered in \cite{Seiberg:1996bd} and continued to be 
studied in~\cite{Morrison:1996xf, Douglas:1996xp,Intriligator:1997pq,Aharony:1997ju,Aharony:1997bh,DeWolfe:1999hj}.
Recently, by using more modern techniques such as localization  computation of partition functions, properties of 5d SCFTs are studied 
in much more detail, see e.g., \cite{Benini:2009gi,Bao:2011rc,Kim:2012gu,Iqbal:2012xm,Bashkirov:2012re,Bergman:2013koa,Bergman:2013ala,Bao:2013pwa,Hayashi:2013qwa,Taki:2013vka,Bergman:2013aca,Aganagic:2014oia,Taki:2014pba,Hwang:2014uwa,Zafrir:2014ywa,Hayashi:2014wfa,Bergman:2014kza,Hayashi:2014hfa,Mitev:2014jza,Kim:2015jba,Hayashi:2015xla}. 

One of the main focus of these works is on enhancement of global symmetry.
Let us see this from a top-down point of view in the sense of renormalization group (RG) flows. 
Suppose that we have a 5d SCFT with a global symmetry $G$.
We can deform the theory by mass parameters.
The theory has 5d $\CN=1$ supersymmetry (i.e., 8 supercharges), and hence its mass deformation $m$ takes values in the
Cartan subalgebra of the flavor symmetry group $G$.
After the mass deformation, conformal invariance is broken and a nontrivial RG flow is triggered. If the mass deformation is generic enough,
we typically get a weakly coupled low energy effective field theory which is a supersymmetric gauge theory.\footnote{
Let us make the point of view a little more clear.
We regard 5d supersymmetric gauge theories as low energy effective field theories of mass-deformed 5d (or sometimes 6d) SCFTs. 
The relation between 5d gauge theories and UV mass-deformed SCFTs is somewhat analogous to 
the relation between the effective Lagrangian of pions and QCD.
For example, the fact that some of the BPS states of the mass-deformed SCFTs are realized as instanton particles in 5d gauge theories 
is analogous to the fact that the pion effective Lagrangian can, surprisingly, realize baryons as skyrmions~\cite{Witten:1983tx}. 
A nontrivial point of 5d supersymmetric gauge theories is that they seem to be able to realize all the BPS states of the UV mass-deformed SCFTs.
This is one of the reasons that 5d gauge theories are often regarded as Lagrangians by dropping the adjective ``effective".
However, we do not expect that 5d gauge theories themselves are well-defined up to arbitrary high energy including non-BPS information,
because of nonrenormalizability. 
See, e.g., \cite{Bern:2012di}.}
Now, if the mass deformation is taken in the Cartan part of a non-abelian flavor symmetry, the symmetry $G$ is broken down to a subgroup $H$
at low energies. This means that the symmetry visible in the low energy effective gauge theory $H$ is smaller than the symmetry $G$
of the UV SCFT. Following the renormalization group flow in the inverse direction, it looks like that
the symmetry $H$ of the low energy gauge theory is enhanced to $G$ in the UV limit.

Therefore, it is important to see the symmetry enhancement in the study of ``inverse RG flows" from low energy gauge theories to UV SCFTs.
Various methods have been used to see the symmetry enhancement in the above references.

Recently, a particularly simple method of determining the symmetry enhancement was proposed \cite{Tachikawa:2015mha}
and further investigated in \cite{Zafrir:2015uaa}.
One of the visible global symmetries of the low energy gauge theory is the one associated to the current $j \sim * \tr F^2$ whose charge
is the instanton number, where $F$ is the field strength of the gauge group.\footnote{Again, this is analogous to the fact~\cite{Witten:1983tx} that
the baryon number in the pion effective Lagrangian is given by the charge associated to the Maurer-Cartan 3-form $\tr (U^{-1} dU)^3$,
where $U$ is the unitary matrix representing pions as Nambu-Goldstone bosons. Pursuing the analogy, Chern-Simons terms of 
5d gauge theories are also analogous to Wess-Zumino-Witten terms in the pion theory.}
This, by itself, is a $\U(1)$ symmetry. Now we can consider instanton operators in five dimensions, which are
defined by the property that the instanton number on a small $S^4$ surrounding this operator is nonzero.
If there is an instanton operator which is also a (broken) current multiplet, this current multiplet is charged under the $\U(1)$.
Then the $\U(1)$ current and the instanton operator must form a non-abelian global symmetry. The operator spectrum of the instanton operators are determined 
just by a simple semi-classical quantization of zero modes around instantons. In this way, we can 
determine the enhanced symmetry of the UV SCFT involving instanton currents.\footnote{Actually, this method is more general than just determining the enhanced symmetry.
We can also find other BPS operators as realized as instanton operators. See \cite{Hayashi:2014hfa,Zafrir:2015uaa}.} 
Furthermore, if the Dynkin diagram of the enhanced symmetry is of affine type, that means
the theory is actually an $S^1$ compactification of a 6d SCFT~\cite{Tachikawa:2015mha}.

Other than its simplicity, the method also has the advantage that it does not require any string theory realization of the theory.
We just need the low energy gauge theory. Of course, we do not know whether a given 
non-renormalizable gauge theory can be UV-completed by an SCFT or not
unless the theory is explicitly realized by string theory. However, we will see that the method of \cite{Tachikawa:2015mha} might even be used 
to see whether a given gauge theory has UV completion or not; if the symmetry enhancement gives an inconsistent (i.e., neither finite nor affine) Dynkin diagram,
that is an indication that the theory does not actually exist.
For example, in the case of $\SU(N)$ gauge theory with $N_f$ flavors,
this argument suggests a bound $N_f \leq 2N+4$ which was also found in \cite{Kim:2015jba} by a different method.

In this paper we study symmetry enhancement in general $\SU(N)$ quiver gauge theories whose nodes form an ADE Dynkin diagram of type $G$.
Each node labeled by $i$ has $\SU(N_i)$ gauge group, Chern-Simons level $\kappa_i$ and fundamental flavors $w_i$ in addition to the bifundamentals
dictated by the quiver diagram. We denote the Cartan matrix of the Dynkin diagram corresponding to the quiver as $A_{ij}$. Then 
the usual combination $2N-N_f$ of the gauge group rank and flavor number at each node $i$ is given by $\sum_j A_{ij} N_j - w_i$.
We would like to summarize main results. 

In sec.~\ref{sec:less}, we study the case
in which $2|\kappa_i| \leq \sum_j A_{ij} N_j - w_i$ is satisfied at each node. The rules for the symmetry enhancement is as follows.
We prepare two copies of the Dynkin diagram of the quiver and denote them as ${\rm Dyn}_+$ and ${\rm Dyn}_-$.
Then we remove the node $i$ from ${\rm Dyn}_+$ (reps. ${\rm Dyn}_-$) if the Chern-Simons level is such that
$+2\kappa_i \neq \sum_j A_{ij} N_j - w_i$ (resp. $-2\kappa_i \neq \sum_j A_{ij} N_j - w_i$). After this process, we get Dynkin diagrams which
define a subalgebra of $G \times G$. 
This subalgebra gives the enhanced symmetry. One instanton operator of $\SU(N_i)$ gives the generator corresponding to the simple root 
at the node $i$ of ${\rm Dyn}_\pm$ when $\pm 2\kappa_i = \sum_j A_{ij} N_j - w_i$.
In particular, if $\kappa_i=0 $ and $\sum_j A_{ij} N_j - w_i=0$
at all the nodes, we get the full $G \times G$ flavor symmetry as already discussed in \cite{Tachikawa:2015mha}.

In sec.~\ref{sec:more}, we study the case in which
the condition $2|\kappa_i| \leq \sum_j A_{ij} N_j - w_i $ is violated at some of the nodes. We can get further enhancement of the symmetry.
In particular, if $\kappa_i=0$ and $\sum_j A_{ij} N_j - w_i=-2$ at a node $i$, the Dynkin diagrams of ${\rm Dyn}_+$, ${\rm Dyn}_-$
and the flavor $\SU(w_i)$ are combined into a larger symmetry. The rule is that the node $i$ of ${\rm Dyn}_+$ and the node $i$ of ${\rm Dyn}_-$ 
at which $\sum_j A_{ij} N_j - w_i=-2$ is satisfied are connected by
the Dynkin diagram of $\SU(w_i)$, and we get a larger Dynkin diagram.
If we get an affine Dynkin diagram, the theory should be interpreted as an $S^1$ compactification of a 6d SCFT, and if we get an inconsistent Dynkin diagram,
we might exclude that theory as mentioned above.

In sec.~\ref{sec:4}, we also study the case in which hypermultiplets in symmetric and anti-symmetric representations are introduced. 
Very similar story happens also in this case.

Due to the lack of conditions similar to the asymptotic free condition in 4d or anomaly free condition in 6d and 4d,
we do not know what theories are allowed in 5d. (See \cite{Bhardwaj:2013qia,Bhardwaj:2015xxa} for this direction in 4d and 6d.)
However, the author believes that the theories studied in this paper cover
all the ``generic" $\SU(N)$ quiver theories in which the rank of the $\SU(N)$ gauge groups are not too small.
The fact that we find symmetry enhancement of finite/affine/inconsistent Dynkin diagrams depending on the flavor numbers supports this idea.
It would be interesting to refine this point further.
It would also be interesting to study the cases of non-generic (i.e., small rank) theories with non-generic representations,
and quiver theories with other gauge groups~\cite{Zafrir:2015uaa}.
Instanton operators have been proposed and studied in \cite{Lambert:2014jna,Rodriguez-Gomez:2015xwa,Cremonesi:2015lsa} 
with a different focus from that of \cite{Tachikawa:2015mha,Zafrir:2015uaa} and this paper,
and it would be interesting to study connections between these works.

\paragraph{Note added:} A paper \cite{Hayashi:2015fsa} appeared on the same day with this paper in arXiv.\footnote{We thank Futoshi Yagi for 
informing us of the day of submission of their paper to arXiv.} There, the theories with $N_f>2N-|\kappa|$ are explicitly constructed by brane constructions, and 
the symmetry enhancement is derived by a different method from that of this paper, with completely the same results.

See also \cite{Gaiotto:2015una} which uses the same instanton method as this paper.

\section{Preliminary}~\label{sec:2}
Let us briefly recall the procedure of \cite{Tachikawa:2015mha}. 
In 5d gauge theories, instanton currents $j  \sim * \tr F^2$ define global $\U(1)$ symmetries whose charges are the instanton numbers. 
We want to see that they are enhanced to non-abelian symmetries whose Cartan part is given by the instanton symmetries (and other $\U(1)$ symmetries).
The non-Cartan part must be charged under the Cartan $\U(1)$ symmetries, and hence they must have nontrivial instanton numbers.
In the UV limit the theory is conformal, so by using the state-operator correspondence, the non-Cartan part is identified with states on $S^4 \times \mathbb{R}$
with nonzero instanton numbers on $S^4$.

In theories with 8 supercharges, a flavor symmetry current is in a supermultiplet whose bottom component
is a space-time scalar operator in the triplet of $\SU(2)_R$ R-symmetry. We want to identify these operators with the states obtained by 
quantizing fermion zero modes of instantons on $S^4$. We will see that one instanton is enough for the purpose of this paper.
Focusing on one instanton, the procedure is as follows (see \cite{Tachikawa:2015mha} for more details).
\begin{enumerate}
\item Consider one instanton configuration on $S^4$ which preserves $\SO(5)$ rotational symmetry of $S^4$.
\item Quantize fermion zero modes of the instanton and obtain states on $S^4$.
\item Pick up states which are $\SO(5)$ singlet, $\SU(2)_R$ triplet and gauge singlet. Identify them as the states corresponding to the lowest components of current supermultiplets of enhanced symmetries.
\end{enumerate}

\subsection{Single $\SU(N)$} \label{sec:pre}
As a preparation of the analysis of more general quivers, we first review the case of a single $\SU(N)$ gauge theory with $N_f$ flavors
and Chern-Simons level $\kappa$. 
We decompose the $\SU(N)$ as $\SU(N) \supset \SU(2) \times \SU(N-2) \times \U(1)$, and embed one instanton into $\SU(2)$.
In the following analysis, we assume that $N \geq 3$ so that nothing non-generic happens.

\paragraph{Gauginos.}
The gauginos are in the adjoint representation, and the adjoint representation is decomposed into
(i) the adjoint of $\SU(2)$, (ii) two bifundamentals of $\SU(2) \times \SU(N-2)$, and (iii) singlets of $\SU(2)$.
All of them are also doublets of $\SU(2)_R$.

From the adjoint of $\SU(2)$, we get complex-4 or real-8 gaugino zero modes. They are in the 4-dimensional representation of $\Sp(2) \cong \SO(5)$
and hence these zero modes are written as $\lambda_{i \alpha}$, where $\alpha=1,2,3,4$ is for $\Sp(2) \cong \SO(5)$ rotational symmetry
and $i=1,2$ is for $\SU(2)_R$. The reality condition $(\lambda_{i\alpha})^*=\epsilon^{ij} J^{\alpha \beta} \lambda_{j \beta}$ is imposed.
The anti-commutation relation is
$
\{\lambda_{i\alpha}, \lambda_{j \beta} \} = \epsilon_{ij} J_{\alpha \beta}.
$
This algebra gives states
\beq
\ket{(ij)},~~~\ket{i\alpha},~~~\ket{[\alpha\beta]}
\eeq
where $\ket{(ij)}$ represents a triplet of $\SU(2)_R$, $\ket{i \alpha}$ is a bifundamental of $\SU(2)_R \times \Sp(2)$, and $\ket{[\alpha\beta]}$
is an anti-symmetric traceless representation of $\Sp(2)$ that is a vector of $\SO(5)$.
They have the correct quantum numbers to be a current supermultiplet.

From the bifundamentals of $\SU(2) \times \SU(N-2)$, we get complex $2(N-2)$ or real $4(N-2)$ zero modes $B_{ia}$ and $(B_{ia})^*$
where $a=1,2,\cdots,N-2$ is for $\SU(N-2)$ and $i=1,2$ is for $\SU(2)_R$ as before. 
The $B_{ia}$ has $\U(1)$ charge $N/(N-2)$ if we normalize the $\U(1)$ generator as 
\beq
Q=\frac{1}{N-2}\diag(N-2,N-2,-2,-2, \cdots, -2).
\eeq
The anti-commutation relation is just
$
\{B_{ia}, (B_{jb})^* \}=\delta_i^j \delta_{a}^b.
$

When the Chern-Simons level is $\kappa$, the one instanton has $\U(1)$ charge $\kappa $ in the above normalization of $\U(1)$.
Taking this into account, the state annihilated by all of $(B_{ia})^*$ has $\U(1)$ charge $(-N+\kappa)$, and we denote this state as
$\ket{(-N+\kappa)}_g$.
Other states are obtained by acting $B_{ia}$ to this state.

We are only interested in states which are $\SU(N-2) \times \U(1)$ singlet, $\SU(2)_R$ triplet
and $\SO(5) \cong \Sp(2)$ singlet. As we will see, zero modes from hypermultiplets in the fundamental representation are always singlet of $\SU(N-2) \times \SU(2)_R \times \SO(5)$.
Thus, we only need to concentrate on the states
\beq
\ket{(ij)} \otimes \ket{(\pm N+\kappa) }_g, \label{eq:gpart}
\eeq
where $\ket{(+N+\kappa) } = (B_{ia})^{2(N-2)} \ket{(-N+\kappa)}_g$.

\paragraph{Fundamental hypers.}
Hypermultiplets are in the fundamental representation of $\SU(N)$, and the fundamental representation is decomposed as a doublet of $\SU(2)$
and singlets of $\SU(2)$. 
From the doublets of $\SU(2)$, we get complex $N_f$ zero modes $C_{s}$ and $(C_{s})^*$, where $s=1,2,\cdots, N_f$ and $\{C_s, (C_t)^* \}=\delta_s^t$.
The $C_s$ has $\U(1)$ charge $1$. There is also a baryon number symmetry acting on the hypermultiplets. 
If we take the hypermultiplet baryon charge to be $1$,
the states are obtained by acting $C_{s}$ to the state $\ket{-N_f/2, -N_f/2}_h$ which has 
$\U(1)$ charge $-N_f/2$ and baryon charge $-N_f/2$,
\beq
&\ket{\wedge^\ell, (-N_f/2+\ell), (-N_f/2+\ell)}_h \nonumber \\
:=& C_{s_1} \cdots C_{s_\ell}\ket{-N_f/2, -N_f/2}_h~~~(\ell=0,1,\cdots,N_f).\label{eq:hpart}
\eeq
where $\wedge^\ell$ means the $\ell$-th anti-symmetric representation of the flavor $\SU(N_f)$ group.

\paragraph{Symmetry enhancement.}
We need to combine \eqref{eq:gpart} and \eqref{eq:hpart} to get $\U(1)$ invariant states.
There are at most two states given by
\beq
\ket{E_+} := &\ket{(ij)} \otimes \ket{(- N+\kappa) }_g \otimes \ket{\wedge^{+ N-\kappa+N_f/2}, ( + N-\kappa), ( + N-\kappa)}_h ~\nonumber \\
& (\text{only if ~} |+ N-\kappa| \leq N_f/2) .  \label{eq:rootp}  \\
\ket{E_-} := &\ket{(ij)} \otimes \ket{(+ N+\kappa) }_g \otimes \ket{\wedge^{- N-\kappa+N_f/2}, ( - N-\kappa), ( - N-\kappa)}_h ~\nonumber \\
& (\text{only if ~} |- N-\kappa| \leq N_f/2) .\label{eq:rootm}
\eeq
For the moment, let us restrict ourselves to the case where the condition $|\kappa|+N_f/2 \leq N$ is satisfied. Then,
combined with $ |\pm N-\kappa| \leq N_f/2$, the only allowed possibilities are
\beq
\ket{E_+} = &\ket{(ij)} \otimes \ket{-N_f/2}_g \otimes \ket{\wedge^{N_f}, N_f/2, N_f/2}_h ~\nonumber \\
& (\text{only if ~} \kappa = N-N_f/2) . \\
\ket{E_-} = &\ket{(ij)} \otimes \ket{N_f/2}_g \otimes \ket{\wedge^{0}, -N_f/2, -N_f/2}_h ~\nonumber \\
& (\text{only if ~} \kappa = -N+N_f/2)) .
\eeq
In this case, $\ket{E_\pm}$ is $\SU(N_f)$ singlet and has baryon charge $\pm N_f/2$.

We denote the instanton number as $I$, the baryon number as $B$. 
Then we define 
\beq
H_{\pm}=\frac{1}{2}I \pm \frac{B+\kappa I}{2N}. \label{eq:su2cartan}
\eeq
The baryon charge of $\ket{E_{\pm}}$ is $B= \pm N- \kappa$, and hence 
the states $\ket{E_{\pm}}$, if they exist, have charges $(H_+, H_-)$ given by $(1,0)$ and $(0,1)$, respectively.
Anti-instanton should give states $\ket{(E_{\pm})^*}$ whose charges are opposite of those of $\ket{E_{\pm}}$.

Therefore the $\U(1)_+ $ symmetry is enhanced to $\SU(2)_+$ if $\kappa = N-N_f/2$, and the same is true for $\U(1)_-$ if $\kappa=-N+N_f/2$.
Note that if $N-N_f/2=0$ and $\kappa=0$, both of $\SU(2)_\pm$ are enhanced. 

What is actually shown by the above argument is that $H_\pm$ is the Cartan part of the $\SU(2)_\pm$
symmetry up to addition of flavor $\U(1)$ charges such that the charges of $\ket{E_\pm} $ remain to be $+1$. 
When $\kappa=0=N-N_f/2$, both the baryon $\U(1)$ and the instanton $\U(1)$ are incorporated into the non-abelian 
symmetry $\SU(2)_+ \times \SU(2)_-$, so there is no
ambiguity if the symmetry is $\SU(2)_+ \times \SU(2)_-$. 
However, when only e.g. $\SU(2)_+$ is enhanced, the Cartan of this $\SU(2)_+$ still has an ambiguity of the form 
$H_+ + c H_-$, where $c$ is the yet undetermined constant. This is because the charge of $\ket{E_+}$ is still $+1$ for arbitrary $c$.
We want to fix this ambiguity. 

To do so, notice that the theory with $\kappa=N-N_f/2$ can be achieved from the theory with $2N$ flavors
by adding $2N-N_f$ flavors the masses of positive sign. (The sign depends on convention. The only important point here is that 
the masses of all the massive quarks have the same sign. If all the signs are reversed, we get a theory with opposite sign of $\kappa$.) 
Integrating out the massive quarks, we get the theory
with $N_f$ flavors and the desired Chern-Simons level $\kappa=N-N_f/2$. Now, integrating out the massive quarks has the 
effect that it produces a Chern-Simons coupling between the baryon flavor gauge field $A_B$ and the $\SU(N)$ gauge field.
It is just a one-loop computation\footnote{A simpler way to check this (see e.g., \cite{Ohmori:2014kda}) is that the zero modes (up to quark masses) of these $n_f=2N-N_f$ massive quarks
around one instanton give states with baryon charges $-n_f/2, -n_f/2+1, \cdots, n_f/2$, and only the state with the charge $-n_f/2$ (which has the
lowest energy) contributes to the low energy theory with $N_f=2N-n_f$ flavors.
} 
and is given as
\beq
-\frac{2N-N_f}{4(2\pi)^2}  \int A_B \tr F^2=\frac{\kappa}{2(2\pi)^2}  \int A_B \tr F^2,\label{eq:BCS}
\eeq
where the trace is in the fundamental representation of the gauge group $\SU(N)$. This means the following.
Let $B_{\rm UV}$ be the baryon symmetry of the theory with $2N$ flavors, and let $B_{\rm IR}$ be the one in the theory with $N_f$ flavors
without the term \eqref{eq:BCS}. Then, they are related as $B_{\rm UV} = B_{\rm IR}+\kappa I$ because, in the instanton background,
the term \eqref{eq:BCS} gives instantons additional baryon charge $ \kappa I$. Now, we have argued above that 
the Cartan of $\SU(2)_+$ is unambiguously given by $H_+ = I /2+B_{\rm UV}/2N$ in the theory with $2N$ flavors.
Presumably this $\SU(2)_+$ is unchanged in the RG flow from the theory with $2N$ flavors to the theory with $N_f$ flavors.
(One might wonder about the $\SU(2)_-$ symmetry which is present in the theory with $2N$ flavors but is absent in the theory with $N_f$
flavors. The point is that the above masses of quarks is not traceless, so it is in the Cartan of the $\U(2N)$ flavor symmetry but not in $\SU(2N)$.
However, the $\U(1)$ part of the $\U(2N)$ flavor symmetry is actually a part of the $\SU(2)_+ \times \SU(2)_-$ symmetry of the UV superconformal theory,
and hence the addition of masses in the $\U(1)$ part must break at least one of the $\SU(2)_+$ or $\SU(2)_-$. We need to tune the mass deformation 
associated to the instanton charge such that
$\SU(2)_+$ is preserved. Then $\SU(2)_-$ must be broken to its Cartan.)

Therefore the Cartan of $\SU(2)_+$ in the $N_f$ flavor theory is given by 
\beq
H_+ = \frac{1}{2} I +\frac{B_{\rm UV}}{2N}= \frac{1}{2}I+\frac{B_{\rm IR}+\kappa I}{2N}.
\eeq
This confirms our claim that the Cartan of $\SU(2)_+$ is really given by $H_+$ in \eqref{eq:su2cartan}.
The same is true for $\SU(2)_-$ when $\kappa=-N+N_f/2$.
Note that gauge invariant baryon chiral operators have charge $H_+ = \pm 1/2$, so they are a doublet of $\SU(2)_+$.

\subsection{$N_f >2N$}\label{sec:largenf}
It is very interesting to consider the case $N_f >2N$.
The restriction $2N-N_f-2|\kappa| \geq 0$ was obtained in \cite{Intriligator:1997pq} under the condition that the whole Coulomb branch can be described
as an $\SU(N)$ gauge theory. However, if we allow the possibility of phase transitions on the Coulomb branch, we may still have theories
with $2N-N_f-2|\kappa| < 0$.

What values of $N_f$ and $\kappa$ can be consistent? 
The instanton operators are supposed to correspond to generators of the symmetry $G_{\rm UV}$ of the UV superconformal theory
which contains $\SU(N_f)$ as a subgroup.
Then, we should never get representations $\wedge^3, \cdots, \wedge^{N_f-3}$ since there is no symmetry breaking $G_{\rm UV} \to \SU(N_f)$
under which the adjoint representation of $G_{\rm UV}$ produces these representations (when $N_f >8)$. Therefore, the parameter region
$ | \pm N - \kappa| \leq N_f/2 -3 $
is excluded by using \eqref{eq:rootp} and \eqref{eq:rootm}. Assuming that the Chern-Simons level is not too large, e.g., $|\kappa| \leq N$,
the excluded region is $N_f \geq 2N-2|\kappa|+6$. However, there is a possibility that theories with 
\beq
N_f \leq 2N-2|\kappa|+4 \label{eq:const}
\eeq
might exist. (Recall that $N_f \equiv 2\kappa$ mod 2 due to parity anomaly, so $N_f=2N-2|\kappa|+5$ cannot occur.)
The existence of such theories is really argumed recently \cite{Kim:2015jba}.

As illustrative examples, let us consider the cases $\kappa=0$, $N_f=2N+2$ or $N_f=2N+4$.
As before, $\ket{E_\pm} $ has  charge $(H_+, H_-)=(1,0)$ and $(0,1)$ respectively, so $(H_\pm, E_\pm, (E_\pm)^*)$
form $\SU(2)_\pm$ sub-algebras. However, they also transform in nontrivial representations of $\SU(N_f)$.
The Dynkin diagram of $\SU(N_f)$ is
\begin{equation}
\node{}-\node{}-\node{}-\cdots -\node{}-\node{}-\node{} 
\end{equation} 
where there are $N_f-1$ nodes. When $N_f=2N+2$, the $E_+$ and $E_-$ transform in the fundamental and anti-fundamental representation
of $\SU(N_f)$, respectively as one can check by \eqref{eq:rootp} and \eqref{eq:rootm}. 
This means that we get a new Dynkin diagram
\beq
\blacknode{}-\node{}-\node{}-\node{}-\cdots -\node{}-\node{}-\node{} - \blacknode{} \label{eq:2n2}
\eeq
where the two black nodes come from the $\SU(2)$ subalgebras of the instanton operators $(H_+, E_+, (E_+)^*)$ and $(H_-, E_-, (E_-)^*)$, respectively.
Therefore the symmetry is enhanced to $\SU(N_f+2)$. 

When $N_f=2N+4$, it was predicted \cite{Kim:2015jba} that the theory has a UV completion as $S^1$ compactification of a 6d SCFT. 
In this case we get an affine Dynkin diagram
\beq
\node{}-\node{\blackver}-\node{}-\cdots -\node{}-\node{\blackver}-\node{} .
\eeq
As discussed in \cite{Tachikawa:2015mha}, when the Dynkin diagram becomes affine, the theory is in fact $S^1$ compactification of a 6d SCFT.
Thus the symmetry enhancement is consistent with the claim of \cite{Kim:2015jba} that this is indeed a 6d SCFT.
The symmetry corresponding to this affine Dynkin diagram is $\SO(2N_f)=\SO(4N+8)$, so this should be the symmetry of the 6d SCFT.

Other cases of $N_f$ and $\kappa$ are listed in the table~\ref{tab:nflarge}.
The argument of this paper alone does not guarantee the existence of these theories, but
at least their Dynkin diagrams are consistent.

\begin{table}[t]
\centering
\begin{tabular}{|c|c|c|c|c|}
\hline
 &$N_f= 2N+1$  &$N_f= 2N+2$  &$N_f= 2N+3$  &$N_f= 2N+4$ \\ \hline
$|\kappa|=0$ & &$\SU(N_f+2)$ && $\widehat{\SO(2N_f)}$ \\ \hline
$|\kappa|=1/2$ &$\SU(N_f+1) \times \SU(2)$&&$\SO(2N_f+2)$& \\ \hline
$|\kappa|=1$ &&$\SO(2N_f) \times \SU(2)$&& \\ \hline 
$|\kappa|=3/2$ & $\SO(2N_f)$ &&& \\ \hline
\end{tabular}
\caption{Symmetry enhancement in $N_f>2N$ theories. The conditions $N_f=2\kappa$ mod~2 and $N_f \leq 2N -2|\kappa|+4$ are imposed.
The hat on $\SO(2N_f)$ means that the corresponding Dynkin diagram is affine.}
\label{tab:nflarge}
\end{table}

All the 6d SCFTs are conjectured to be realized by F-theory on elliptic Calabi-Yau threefolds 
\cite{Heckman:2013pva, DelZotto:2014hpa,Heckman:2015bfa,Heckman:2015ola}. 
There is actually a candidate 
for the above 6d theory. In the terminology of those papers, it is a theory supported on a single curve of self-intersection $-1$ with $\Sp(N-2)$ gauge group
which is coupled to $\frac{1}{2}(4N+8)$ fundamental hypermultiplets.
It is also the same as the minimal $(\SO(2N+4),\SO(2N+4))$ conformal matter~\cite{DelZotto:2014hpa}.
The global symmetry of this theory is $\SO(4N+8)$, and the dimension of the Coulomb branch, after compactification to 5 dimensions, is 
$+1+(N-2)=N-1$ where the first $+1$ comes from a tensor multiplet in six dimensions. This is the same as the rank of $\SU(N)$.
It would be interesting to check this conjecture in more detail.
Note that when $N=2$, the conjecture reduces to the well-known fact about rank 1 E-string theory.

\section{Symmetry enhancement in general quivers}\label{sec:3}
Let us consider 5d quiver gauge theories whose nodes form a simply laced (possibly affine) Dynkin diagram of type $G$ with Cartan matrix $A_{ij}$.
Gauge group at the node labelled by an index $i$ is $\SU(N_i)$, and we consider the case that this is coupled to $w_i$ additional flavors in the fundamental
representation and has Chern-Simons level $\kappa_i$. 
We want to study the symmetry enhancement of this class of theories.

\subsection{$N_f \leq 2N$ and $G \times G$ symmetry enhancement}\label{sec:less}
In this subsection, we study the case that the parameters of the theory satisfy
\beq
\sum_j A_{ij} N_j -w_i-2|\kappa_i| \geq 0 ,\label{eq:flavorcondition}
\eeq
where $A_{ij}$ is the Cartan matrix associated to the ADE Dynkin diagram defined as $A_{ii}=2$, $A_{ij} =-1$ if $i$ and $j$ are adjacent, and zero otherwise. 
There are of course bifundamentals between the nodes $i$ and $j$ if $A_{ij}=-1$. The above condition means that the total flavor number at each node
is not larger than $2N_i -2|\kappa_i|$.
We also assume that the ranks of the gauge groups satisfy $N_i \geq 3$.
The case with $\SU(2)$ gauge groups will be treated in sec.~\ref{sec:su2}.
In this class of theories, we want to identify enhanced symmetries
in the UV SCFT in addition to the visible symmetries.

First, we need to define baryon symmetries appropriately. 
Let us focus on the node labelled by $i$. We define a baryon symmetry $B_i$ associated to this node as follows.
The $B_i$ baryon charges of $w_j$ fundamentals are defined to be $ \delta_{ij}$. Next, let us consider a bifundamental $Q_{ij}$
of $\SU(N_i) \times \SU(N_j)$ for $A_{ij} = -1$. We use a convention that $Q_{ij}$ is in the fundamental representation of $\SU(N_i)$ and 
in the anti-fundamental representation of $\SU(N_j)$. Then $Q_{ji}$ is the CPT conjugate of $Q_{ij}$.
The charge of $Q_{ij}$ under $B_i$ is defined to be $1$. Note that in this convention, the charge of $Q_{ij}$ under $B_j$
is $- 1$ because of the CPT conjugation. Thus, the charge under $B_i$ is $+1$ if the field is in the fundamental representation of $\SU(N_i)$,
$-1$ if it is in the anti-fundamental representation of $\SU(N_i)$, and zero otherwise.

We denote the instanton number associated to the gauge group $\SU(N_i)$ as $I_i$. 
Define
\beq
H_{i,\pm}=\left(\frac{1}{4} \sum_j A_{ij} I_j \right)\pm \left(\frac{B_i + \kappa_i  I_i}{2N_i } \right).\label{eq:cartandef}
\eeq
When $I_i=1$ and $I_j=0$ ($j \neq i$), we get
\beq
H_{i,\pm} &=\frac{1}{2} \pm \frac{B_i+\kappa_i}{2N_i},  \label{eq:daigcharge} \\
H_{j,\pm} &=-\frac{1}{4} \pm \frac{B_j}{2N_j} ~~~(A_{ij}=-1) \label{eq:nondiagcharge}\\
H_{k,\pm} &= \pm \frac{B_k}{2N_k} ~~~~~~~~~(A_{ik}=0)
\eeq

Now let us consider one instanton of the gauge group $\SU(N_i)$.
All the fields charged under this gauge group has $B_i$ charge $1$,
if these fields are considered to be in the fundamental (i.e., not anti-fundamental) of $\SU(N_i)$ by using appropriate CPT conjugation if necessary.
Therefore, by using the result of the single $\SU(N)$ discussed above, we get states
\beq
&\ket{E_{i, +}}~~~\text{if~~}2\kappa_i= +(\sum_j A_{ij} N_i -w_i ),\\
&\ket{E_{i, -}}~~~\text{if~~}2\kappa_i= -(\sum_j A_{ij} N_i -w_i) .
\eeq
These states $\ket{E_{i,\pm}}$ have the baryon charge $\pm N_i - \kappa_i$ as seen in the previous section,
and hence the charges of these states under $(H_{i,+},H_{i,-} ) $ is given by $(1,0)$ and $(0,1)$ respectively as can be seen from \eqref{eq:daigcharge}.
In fact, the result of the single $\SU(N)$ case implies that $H_{i, \pm}$, $E_{i, \pm}$ and $(E_{i, \pm})^*$ satisfy $\SU(2)$ algebra.

Strictly speaking, the result of the previous section only shows that $H_{i, \pm}$ is the Cartan part of these $\SU(2)$ algebras up to corrections
of the form $\sum_{j \neq i} (c_j \pm \kappa_j d_j)   I_j$ where $c_j$ and $d_j$ are constants, 
because we are setting $I_j=0$ for $j \neq i$ in the analysis. The specific form $c_j \pm \kappa_j d_j$ is dictated by a
consideration of parity and charge conjugation (spurious) symmetries. 
When $\kappa_i=0$ and $\sum_j A_{ij} N_j -w_i=0$ at each node, the only consistent possibility seems to be
that \eqref{eq:cartandef} is actually the Cartan generators. 
More general values of $\kappa_i$ and $w_i$ may be argued along the lines of the argument given in sec.~\ref{sec:pre}
by adding masses to some of the $w_i$ flavors.

Now, the crucial point is that $\ket{E_{i,\pm}}$ are charged under $H_{j,\pm}$ for $j$ such that $A_{ij}=-1$.
Remember that the bifundamental $Q_{ij}$ has $B_j$ charge $- 1$.
Noting that there are $N_j$ zero modes coming from this bifundamental $Q_{ij}$ in the $\SU(N_i)$ instanton, the $B_j$ charge of $\ket{E_{i, \pm}}$
is given as $ \mp N_j/2$. Using \eqref{eq:nondiagcharge}, we get the charges of $\ket{E_{i, +}}$ and $\ket{E_{i, -}}$
as $(H_{j,+}, H_{j,-})=(-1/2, 0)$ and $(0, -1/2)$, respectively.
It is also obvious that we get $H_{k,\pm} =0 $ for $k$ such that $A_{ik}=0$.

Summarizing the above results, the charges of $\ket{E^i_\pm}$ are given as
\beq
H_{j,\pm} \ket{E_{i,\pm}} = \frac{1}{2} A_{ij} \ket{E_{i,\pm}},~~~H_{j,\mp} \ket{E_{i,\pm}} = 0.
\eeq
Now we can clearly recognize that the charges of $\ket{E_{i,\pm}}$  under $H_{j,\pm}$ are exactly the same as that of the generator corresponding to 
the simple root of the node $i$ of the Dynkin diagram. 
We conclude that {\it one-instanton operators at the node $i$ give the generator corresponding to the simple root of $G_+$ ($G_-$)
if the Chern-Simons level satisfies $2\kappa_i= (\sum_j A_{ij} N_i -w_i )$ ($2\kappa_i= -(\sum_j A_{ij} N_i -w_i )$)}.
We have not analyzed multi-instanton operators, but the existence of (some of) the other roots are ensured once we have (some of) the simple roots.

We summarize the rules of the symmetry enhancement.
\begin{enumerate}
\item Take two copies of the Dynkin diagram of type $G$ which we denote as ${\rm Dyn}_\pm$.
\item At a node $i$, if the Chern-Simons level is such that $2\kappa_i \neq (\sum_j A_{ij} N_j -w_i )$, we eliminate
the node $i$ from ${\rm Dyn}_+$.
The same is true for ${\rm Dyn}_-$ if $2\kappa_i \neq -(\sum_j A_{ij} N_j -w_i )$.
\item The remaining Dynkin diagrams ${\rm Dyn}_\pm$ after the above process define some semi-simple Lie algebra which is a subgroup of $G_+ \times G_-$. 
This is the non-abelian part of the enhanced symmetry.
\end{enumerate}
Our claim is that there is at least the symmetry enhancement described above. It can happen that we obtain a larger symmetry
which contains the above one as a subgroup. For example, such further symmetry enhancement can happens when there are $\SU(2)$ gauge groups.

A particularly interesting case is when $\kappa_i=0$ and $\sum_j A_{ij} N_j -w_i =0$. Then we get the symmetry $G_+ \times G_-$.
This case was already discussed in \cite{Tachikawa:2015mha}.

\subsection{$N_f >2N$ and a large class of possible 6d SCFTs} \label{sec:more}
In this subsection, we would like to study the cases in which the condition \eqref{eq:flavorcondition} is violated at some of the nodes.
We will see that the two Dynkin diagrams ${\rm Dyn}_\pm$ discussed in the previous subsection are connected at those nodes
and we get a larger Dynkin diagram.
In many cases the theory is lifted to a 6d SCFT.
We still assume $N_i \geq 3$.

Suppose that we have a quiver of the form of Dynkin diagram of type $G$.
At most of the nodes, we take $w_i$ to satisfy \eqref{eq:flavorcondition}. Then we get the Dynkin diagrams ${\rm Dyn}_\pm$ as discussed before.

Now we choose a node $ \ell$, and take the flavor number and the Chern-Simons level of that node to be 
$w_\ell =2+\sum_j A_{\ell j} N_j$ and $\kappa_\ell=0$, in which case the most interesting thing happens.
As discussed in sec.~\ref{sec:largenf}, the instanton operators $\ket{E_{\ell, \pm}}$ are in a nontrivial representation of the flavor group.
Focusing on gauge invariant states, $\ket{E_{\ell, +}}$ is in the fundamental representation of $\SU(w_\ell)$, and 
$\ket{E_{\ell, -}}$ is in the anti-fundamental representation of $\SU(w_\ell)$.
This means the following. We have two Dynkin diagrams ${\rm Dyn}_\pm$ as in the previous subsection.
However, now {\it the nodes $\ell$ of ${\rm Dyn}_+$ and ${\rm Dyn}_-$ are connected by the Dynkin diagram of $\SU(w_\ell)$.}

To illustrate the point, let us focus on the case where $\kappa_i=0$ and $w_i =\sum_j A_{i j} N_j$ at most of the nodes other than $\ell$ where 
$w_\ell =2+\sum_j A_{\ell j} N_j$. Furthermore, we take the node $\ell$ to be at the end of the Dynkin diagram.
Then we get the following Dynkin diagrams.

When the quiver is of $A_{n}$ type, the symmetry enhancement gives 
\beq
\blacknode{}-\cdots - \blacknode{} - \node{}- \cdots - \node{} - \blacknode{}- \cdots - \blacknode{},
\eeq
where black nodes are from instanton operators and white nodes are from $\SU(w_\ell)$.
Note that the two Dynkin diagrams of the black nodes are connected by the Dynkin diagram of $\SU(w_\ell)$ which consists of the white nodes.
Thus we get $\SU(2n+w_\ell)$.

When the quiver is of $D_{n}$ ($n \geq 3$) type, the symmetry enhancement gives
\beq
\blacknode{}-\blacknode{\blackver}-\cdots - \blacknode{} - \node{}- \cdots - \node{} - \blacknode{}- \cdots -\blacknode{\blackver}- \blacknode{},
\eeq
This is affine $D_{2n+w_\ell-2}$. So the theory is actually lifted to a 6d SCFT with $\SO(2(2n+w_\ell-2))$ flavor symmetry.

When the quiver is of $E_{6,7,8}$ type, the symmetry enhancement gives
\beq
\blacknode{}-\blacknode{}-\blacknode{\blackver}-\cdots - \blacknode{} - \node{}- \cdots - \node{} - \blacknode{}- \cdots -\blacknode{\blackver}- \blacknode{}-\blacknode{}.
\eeq
There are no such (affine) Dynkin diagrams, so probably the theory cannot be UV completed.

There is another way to get affine Dynkin diagrams. Take $A_{n}$ type quiver and let $\ell$ and $\ell'$ be the nodes at the two ends of the $A_n$ quiver.
If we take $w_\ell =2+\sum_j A_{\ell j} N_j$ and $w_{\ell'} =2+\sum_j A_{\ell' j} N_j$ at these two nodes, we get an affine $A_{2n+w_\ell+w_{\ell'}-3}$
diagram
\beq
-\blacknode{}-\cdots - \blacknode{} - \node{}- \cdots - \node{} - \blacknode{}- \cdots - \blacknode{}- \node{}- \cdots - \node{} - ~~~~\text{(two ends connected)} \label{eq:affineA}
\eeq
where the first and the last nodes are supposed to be connected. So this is a 6d theory with $\SU(2n+w_\ell+w_{\ell'}-2)$ flavor symmetry.

In the above, we have focused on the case that most of the nodes have $\kappa_i=0$ and $w_i =\sum_j A_{i j} N_j$.
However, by choosing $\kappa_i$ and $w_i$ appropriately, we can get arbitrary pairs of Dynkin diagrams $({\rm Dyn}_+, {\rm Dyn}_-)$
which is a subset of two copies of the original Dynkin diagram. By connecting these ${\rm Dyn}_+$ and ${\rm Dyn}_-$ as described above, we get 
many more possibilities for consistent Dynkin diagrams.

We do not know whether theories which give consistent (affine) Dynkin diagrams really have UV completion or not. 
Assuming many of them do exist, they might give a large class of 6d theories in the following way.
For each set of integers $\{N_i \}$ such that $\sum_j A_{ij} N_j \geq 0$, we can tune $w_i$ and $\kappa_i$ appropriately to get 
affine Dynkin diagrams.
So we get a 6d SCFT (or several 6d SCFTs) for each such set $\{N_i \}$.
Then we face a big challenge. In \cite{Heckman:2013pva, DelZotto:2014hpa,Heckman:2015bfa,Heckman:2015ola} it was conjectured that 
all 6d SCFTs are realized by F-theory. If that is the case, the theories found here must be identified with 
some of the theories classified there.

Probably a little more safe (but still not completely rigorously proved) thing to say is that we exclude theories which give an inconsistent Dynkin diagram.
Basically, we need to satisfy $w_i+2|\kappa_i| \leq \sum_j A_{ij} N_j $ at most of the nodes, and only a few nodes can violate
this condition by only a few extra flavors. This is reminiscent of the asymptotic free/conformal condition in 4d $\CN=2$
theories.

\paragraph{Example.} As an example, let us consider $A_n$ type quiver 
\beq
[w_\ell=N+2]-\SU(N)- \cdots- \SU(N)-[w_{\ell'}= N+2],\label{eq:nazoquiver1}
\eeq
where the two ends have $w=N+2$ flavors. This is the type discussed around \eqref{eq:affineA}.
The enhanced symmetry is $\SU(2n+2N+2)$.
 The dimension of the Coulomb branch is $n(N-1)$. 
 
 When $n=2m+1$ ($m=0,1,2, \cdots$) and $N \geq 2m+2=n+1$, 
a candidate for this theory is the one realized in F-theory as
 \beq
\begin{array}{ccccccccc}
 1 &2&2&\cdots&2& \\
 \sp_{N-2m-2} & \su_{2N-4m+4}& \su_{2N-4m+12}& \cdots& \su_{2N+4m-4}& [\su_{2N+4m+4}]
\end{array}\label{eq:nazoquiver2}
\eeq
where the notation means that the $\sp$ is supported on a $-1$ curve, 
$\su$'s are supported on $-2$ curves, and the last $[\su_{2N+4m+4}]$ is the flavor symmetry
supported on a noncompact curve. This type of theories is first constructed by brane setups in \cite{Hanany:1997gh}.
This is an anomaly free valid F-theory model (see section~6 of \cite{Heckman:2015bfa}), 
and has the flavor $\SU(2N+4m+4)=\SU(2N+2n+2)$ symmetry.
The dimension of the Coulomb branch after compactification to 5d is given by
\beq
r&=1+(N-2m-2)+\sum_{p=1}^{m} (1+(2N-4m-4+8p-1 )) \nonumber \\
&=(2m+1)(N-1)=n(N-1),
\eeq
where tensor multiplet contributions are also taken into account.
This dimension is the same as the above quiver \eqref{eq:nazoquiver1}. This may be a natural generalization
of the conjecture discussed in sec.~\ref{sec:largenf} about the $\SU(N)$ $N_f=2N+4$ theory.

In the same way, when $n=2m$ ($m=1,2,\cdots$) and $N \geq 2m=n$, a candidate for the F-theory model is
 \beq
\begin{array}{ccccccccc}
 1 &2&\cdots&2& \\
 \su_{2N-4m+2}& \su_{2N-4m+10}& \cdots& \su_{2N+4m-6}& [\su_{2N+4m+2}].
\end{array}\label{eq:nazoquiver2}
\eeq
The $ \su_{2N-4m+2}$ is supported on a $-1$ curve and hence must be coupled to a hypermultiplet in the 
anti-symmetric representation (see \cite{Hanany:1997gh} and section~6 of \cite{Heckman:2015bfa}).
This has the $\SU(2N+4m+2)=\SU(2N+2n+2)$ flavor symmetry, and the Coulomb branch dimension is $r=2m(N-1)=n(N-1)$.

The above guess is valid only if $N \geq n+1$ (for odd $n$) and $N \geq n$ (for even $n$). What about the opposite case?
In this case, we propose that there is a duality $N-1 \leftrightarrow n$, in the sense that two 5d theories flows from the same UV 6d theory
with different deformation by flavor Wilson loops on $S^1$. 
At the most naive level, this duality may be seen by the brane web construction~\cite{Aharony:1997bh}, and more precise argument might be given
along the lines of \cite{Taki:2014pba,Kim:2015jba}.
Note that the flavor symmetry $\SU(2N+2n+2)$ and the Coulomb branch dimension $r=n(N-1)$ are invariant under the change of $N-1$ and $n$.
By using this duality, the case of smaller $N$ can also be described by the F-theory models described above.
Note that when $N=2$, the correspondence of the F-theory model and the 5d quiver 
is the novel 5d duality of \cite{DelZotto:2014hpa} for the $(\SO(2n+6), \SO(2n+6))$ minimal conformal matter.
Note also that even though there are dualities of 5d gauge theories, the 6d tensor branch has no such duality as discussed in \cite{Heckman:2015ola}.

\subsection{Inclusion of $\SU(2)$ gauge group}\label{sec:su2}
The gauge group $\SU(2)$ is special in several points. First, it does not have Chern-Simons level. Instead it has a discrete theta angle\footnote{
Actually, the discrete theta angle may be regarded as Chern-Simons level in the following sense. Consider an $\SU(2)$ gauge theory in a
five dimensional space $X$. If the gauge configuration is such that the transition function defines a nontrivial element of $\pi_4(\SU(2))={\mathbb Z}_2$,
it cannot be extended to a six dimensional manifold $Y$ whose boundary is $X$, $\partial Y=X$. However,
we can consider $\SU(2)$ bundle as a subbundle of $\SU(3)$. Then, because $\pi_4(\SU(3))=0$, the gauge field on $X$
may be extended to $Y$ as an $\SU(3)$ bundle. In this case we can define Chern-Simons invariant as $\frac{1}{3! (2\pi)^3} \int_Y \tr F^3$.
It is actually a topological invariant as long as the gauge field on the boundary $X=\partial Y$ is restricted to be in $\SU(2)$.
In the case of $X=S^4 \times S^1$ and $Y=S^4 \times D_2$ where $D_2$ is a 2-dimensional disk such that $\partial D_2=S^1$, the 
relevant computation is essentially performed in \cite{Witten:1983tx}, although the computation there is done in the context of
the pion effective theory with a Wess-Zumino-Witten term. 
Let $I$ be the instanton number on $S^4$.
We twist the gauge field by the center of $\SU(2)$ around $S^1$ as $A \to h(t)^{-1}Ah(t)+h(t)^{-1}dh(t)$ where $h(t=0)=1$ and $h(t=2\pi)=-1$,
in such a way that the transition function between the north and south hemispheres of $S^4$ is given as $\hat{g}(y,t)=h(t)^{-1}g(y) h(t)$, where $y$
is the coordinate of the equator of $S^4$ and $g(y)$ is the usual transition function for the instanton on $S^4$.
Note that the gauge field and the transition function are periodic in $t \in S^1$, even though $h(t)$ is not. In this case
the Chern-Simons invariant defined above is $I/2$ mod ${\mathbb Z}$. Therefore the discrete theta angle $\theta$ may be regarded as Chern-Simons
level $\kappa$ as
$\theta/ \pi=\kappa$ mod $2{\mathbb Z}$.}
 $\theta=0, \pi$
associated with $\pi_4(\SU(2)) \cong {\mathbb Z}_2$. Second, in the instanton analysis, it does not have unbroken gauge symmetries
other than the center of $\SU(2)$, $-1 \in \SU(2)$.
Then all the states \eqref{eq:hpart} which are invariant under the center of $\SU(2)$ (including the effect of the discrete theta angle) 
are allowed.
Finally, the fundamental representation of $\SU(2)$ is pseudo-real and hence $N_f$ flavors of quarks have $\SO(2N_f)$ flavor symmetry.
In particular, the baryon symmetry is actually a part of the non-abelian symmetry; $\U(1)_B \times \SU(N_f) \subset \SO(2N_f)$.

In this subsection we want to see what changes are necessary to the results of the previous sections
when the $\SU(2)$ is included in a large generic quiver which contains gauge groups $\SU(N_i)$ with $N_i \geq 3$.

When the $\SU(2)$ gauge group appears in a quiver along with $\SU(N_i \geq 3)$ gauge groups, it is located at the end of the quiver as
\beq
[w]-\SU(2) - \SU(M)- \cdots
\eeq
where $[w]$ means that the $\SU(2)$ is coupled to $w$ fundamentals. If the condition $\sum_j A_{ij} N_j \geq 0$ is satisfied,
the group $\SU(M)$ has $3 \leq M \leq 4$ unless all the gauge groups in the quiver are $\SU(2)$. 
We restrict our attention to this case.

Consider one instanton of $\SU(2)$. For the moment, we assume that $w>0$, and denote the spinor representations of $\SO(2w)$ as $S_\pm$.
If we neglect the conditions imposed by the invariance under $-1 \in \SU(2)$, we get  gauge invariant states 
\beq
\ket{-M/2, S_+},~~~\ket{M/2, S_+},~~~\ket{-M/2, S_-},~~~\ket{M/2, S_-}
\eeq
where $\mp M/2$ is the baryon charge associated to the $\SU(M)$ gauge group.

We need to impose the condition under $-1 \in \SU(2)$. Under the assumption $w>0$,
the parity of ${\rm O}(2w)$ can exchange the two values of the discrete theta angle, so we just set $\theta=0$ without loss of generality.
Then the surviving states are the ones on which the center of $\SU(2)$ gauge group acts trivially.
Depending on whether $M$ is odd or even, the surviving state is
\beq
&\ket{ \mp M/2, S_\pm}~~~(M=\text{odd}),\\
&\ket{\mp M/2, S_+}~~~(M=\text{even}).
\eeq
Now let us study it case-by-case.

When $M$ is odd and $w=1$, we get two states $\ket{ \mp M/2,  \pm 1/2}$, where $\pm 1/2$ is the $\U(1) \cong \SO(2)$ charge. 
In this case, the $\SU(2)$ instanton gives two subalgebras $\SU(2)_+$ and $\SU(2)_-$ which are part of ${\rm Dyn}_+$ and ${\rm Dyn}_-$,
respectively. When $M=3$, the $\SU(2)$ node has $N_f=M+w=4$ and the result here is exactly the same as in sec.~\ref{sec:less}.

When $M$ is odd and $w=2$, the flavor group is $\SO(4) \cong \SU(2) \times \SU(2)$ and
we get two states $\ket{ \mp M/2, S_\pm}$ where $S_+$ is the doublet representation of the first $\SU(2)$ and $S_-$
is that of the second $\SU(2)$. Then, one can see that the instanton operators mix with $\SO(4) \cong \SU(2) \times \SU(2)$
to give subalgebras $\SU(3)_+ \times \SU(3)_-$. These subalgebras $\SU(3)_+ $ and $ \SU(3)_-$ are further connected to the rest of
${\rm Dyn}_+$ and ${\rm Dyn}_-$, respectively. The result here is simply summarized as follows. We formally consider a quiver
\beq
\SU(1)-\SU(2)-\SU(M=3)-\cdots.
\eeq
Then we formally consider ``$\SU(1)$ instanton" and naively apply the result of sec.~\ref{sec:less}.
The ``$\SU(1)$ instanton" plays the role of supplying one doublet of $\SU(2)$ in addition to the bifundamental of $\SU(1)-\SU(2)$.
This was already observed in e.g., \cite{Bergman:2014kza,Hayashi:2014hfa,Tachikawa:2015mha}.

When $M$ is odd and $w=3$, the flavor group is $\SO(6) \cong \SU(4)$ and we get two states $\ket{ \mp M/2, S_\pm}$ where $S_+$ 
and $S_-$ are the fundamental and anti-funamental representations of $\SU(4)$, respectively. 
When $M=3$, the $\SU(2)$ node has $N_f=6$ flavors. Here we encounter a slight modification of the result of sec.~\ref{sec:more}.
In that section, the Dynkin diagrams ${\rm Dyn}_+$ and ${\rm Dyn}_-$ are connected by the Dynkin diagram of $\SU(w=3)$.
However, here they are connected by $\SU(4) \cong \SO(6)$. We get one additional node.

The case $M$ is odd and $w \geq 4$ can be analyzed in a similar way. However, generically they give an inconsistent Dynkin diagram
when ${\rm Dyn}_\pm$ are large enough.

Let us turn to the case of even $M (=4)$. 
When $M$ is even and $w=1$, we get two states $\ket{ \mp M/2,  1/2}$, where $1/2$ is the $\U(1) \cong \SO(2)$ charge.
In this case we seem to still get ${\rm Dyn}_\pm$, but the relation between Cartan subalgebras and the $\U(1)$ charges 
should be changed. We do not work out the detail.

When $M$ is even and $w=2$, the flavor group is $\SO(4) \cong \SU(2) \times \SU(2)$ and
we get two states $\ket{ \mp M/2, S_+}$ where $S_+$ is the doublet representation of the first $\SU(2)$.
The two Dynkin diagrams ${\rm Dyn}_\pm$ are connected by the Dynkin diagram of the first $\SU(2)$ in $\SO(4) \cong \SU(2) \times \SU(2)$.
When $M=4$, the $\SU(2)$ node has the flavor number $N_f=6$. Thus, the result here is the same as in sec.~\ref{sec:more},
except the fact that the $\U(1)$ part of $\U(w=2)=\SU(2) \times \U(1)$ is actually the Cartan of the second $\SU(2)$ in $\SO(4)$.

The case $M$ is even and $w \geq 3$ can be analyzed in a similar way. However, generically they give an inconsistent Dynkin diagram
when ${\rm Dyn}_\pm$ are large enough.

When $w=0$, the result will depend on the value of the discrete theta angle. 
When $M$ is even and $\theta=0$, the nodes in ${\rm Dyn}_\pm$ corresponding to the $\SU(2)$ gauge node survive,
but when $\theta=\pi$, both of them are eliminated. This result is consistent with the results of sec.~\ref{sec:less} if we replace
$\kappa \to \theta/\pi$ mod $2$ (see also the above footnote in this subsection.)
When $M$ is odd, the corresponding node of one of ${\rm Dyn}_+$ or ${\rm Dyn}_-$
is eliminated depending on whether $\theta=0$ or $\theta=\pi$.
This is again consistent with the results of sec.~\ref{sec:less} by $\kappa \to \theta/ \pi - 1/2$ mod $2$.

We also note that when all the nodes are $\SU(2)$, more drastic change can happen.
For example, the theory \eqref{eq:nazoquiver1} actually has affine $\SO(2(2n+6))$ symmetry when $N=2$ instead of $\SU(2n+6) \subset \SO(2(2n+6))$.
Thus the corresponding 6d theory has the $\SO(4n+12)$ symmetry. 
This can be analyzed in the same way \cite{Tachikawa:2015mha}. We leave the detailed analysis to the reader.

\subsection{Mass deformation of SCFT and gauge couplings}
In sec.~\ref{sec:less}, we obtained the relation between instanton numbers $I_i$, baryon numbers $B_i$ and the 
Cartan generators $H_{i, \pm}$ of the symmetry group $G_{\pm}$. This gives us the relation between mass deformation of
$G_+ \times G_-$ in the UV SCFT and the low energy gauge couplings of the gauge groups $\SU(N_i)$.

Let us focus our attention to the case that $\kappa_i=0$ and $\sum_j A_{ij} N_j -w_i=0$ at each node so that the full $G \times G$ symmetries are enhanced.
In this case we have
\beq
H_{i,\pm}=\frac{1}{4} \sum_j A_{ij} I_j \pm \frac{B_i }{2N_i } .
\eeq
Now, in the UV SCFT, let us add mass terms in the Cartan of $G_+ \times G_-$ as
$
\sum_i 2(m_{i, +} H_{i,+} +m_{i, +} H_{i,+} )
$, where the factor $2$ is introduced just for convenience.  From the above equation, we get
\beq
\sum_i 2(m_{i, +} H_{i,+} +m_{i, +} H_{i,+} )= \sum_i (\frac{8\pi^2}{g^2_i} I_i+ m_{B,i}B_i),
\eeq
where
\beq
\frac{8\pi^2}{g^2_i}= \sum_j \frac{1}{2} A_{ij} (m_{i, +} +m_{i, -} ),~~~~~m_{B,i}=\frac{m_{i,+}-m_{i,-}}{N_i}
\eeq
This $8 \pi^2/g^2_i$ is the gauge coupling constant of the low energy gauge theory.

For example, let us consider the case of $A_{n-1}$ quiver 
\beq
\SU(N_1)-\SU(N_2)-\cdots-\SU(N_{n-1}),
\eeq
which can be constructed by brane webs.
In this case, we can take the Cartan generators as $H_{i,\pm}=\frac{1}{2} \diag(0,\cdots,0,1,-1,0,\cdots,0)$ where $i$-th and $i+1$-th components are
$1$ and $-1$, respectively. If we define $\sum_i 2m_{i, \pm} H_{i,\pm}={\mathfrak m}_{\pm}=\diag ({\mathfrak m}_{1,\pm}, \cdots, {\mathfrak m}_{n,\pm})$,
we get ${\mathfrak m}_{i,\pm}-{\mathfrak m}_{i+1,\pm}=\sum_{j}A_{i,j} m_{j, \pm}$ where we used the formula $4\tr (H_{i,\pm} H_{j, \pm})=A_{ij} $.
Combining this with the above result, we get
\beq
{\mathfrak m}_{i,\pm}-{\mathfrak m}_{i+1,\pm}= \frac{8\pi^2}{g^2_i} \pm  \left(\frac{N_{i+1}}{2} (m_{B,i}-m_{B,i+1})-\frac{N_{i-1}}{2} (m_{B,i-1}-m_{B,i}) 
+\frac{w_i}{2} m_{B,i} \right)
\eeq
where $w_i =2N_i-N_{i+1}-N_{i-1}$, and $N_0=N_n=0$ and $m_{B,0}=m_{B,n}=0$. 
Note that $(m_{B,i}-m_{B,i+1})$ is the mass of the bifundamental of $\SU(N_i)-\SU(N_{i+1})$, and $m_{B,i}$
is the mass of the $w_i$ fundamentals at $\SU(N_i)$.
For example, in the case of the 5d $T_N$ theory \cite{Benini:2009gi,Bao:2013pwa,Hayashi:2013qwa,Aganagic:2014oia,Hayashi:2014wfa,Bergman:2014kza,Hayashi:2014hfa,Hayashi:2015xla} the gauge groups are $N_i=i~(i=1,2,\cdots,N-1)$ where $\SU(1)$ is interpreted as 
described in sec.~\ref{sec:su2}. Then the formulas for ${\mathfrak m}_{i,\pm}-{\mathfrak m}_{i+1,\pm}$ above are precisely equivalent to the ones
written down in \cite{Hayashi:2014hfa} up to sign convention and the treatment of $\SU(1)$ explained there.

More generally, if we denote the masses as elements of the Lie algebra of $ G$ as ${\mathfrak m}_\pm$,
the gauge couplings are given as 
\beq
\frac{8\pi^2}{g^2_i} = \tr (({\mathfrak m}_++{\mathfrak m}_-) H_{i}),
\eeq
where $H_i$ are the Cartan generators of $G$, $\tr$ is the inner product on the space of the Lie algebra of $G$ normalized such that 
$4\tr (H_{i} H_{j})=A_{ij} $ holds, and we used the isomorphism $G_+ \cong G_- \cong G$. 
This has an application to 6d SCFTs \cite{Ohmori:2015pia}.

\section{Other matters}\label{sec:4}
In the previous sections, we have only discussed the case that matter hypermultiplets are in (bi)fundamental representations.
Here we consider a hypermultiplet in other representations.
For large enough $N$, presumably the only allowed possibilities are anti-symmetric representation, symmetric representation and adjoint representation.
The adjoint representation was already discussed in \cite{Tachikawa:2015mha}, so we focus on the other cases.
We will see that $\SU(2)_+ $ and $ \SU(2)_-$ are combined into a larger group $\SU(3)$; in other words, 
the nodes of $\SU(2)_+$ and $\SU(2)_-$ are connected.\footnote{After the submission of the first version of this paper, the results of this section have been confirmed
explicitly by brane web construction~\cite{Bergman:2015dpa}.}

\subsection{Anti-symmetric representation}
Let us consider the anti-symmetric representation of the gauge group.
We assume $N>3$, since the anti-symmetric representation for $N=3$ is just an anti-fundamental representation.
For the moment, we also assume $N>4$, since the anti-symmetric representation in the $N=4$ case is a real representation
and hence an additional enhancement occurs.

First, we consider the case of a single $\SU(N)$ gauge group with one anti-symmetric representation and $N_f$ fundamental representations.
Under the decomposition $\SU(N) \supset \SU(2) \times \SU(N-2) \times \U(1)$, the anti-symmetric representation
is decomposed into (i) a bifundametal of $\SU(2) \times \SU(N-2)$ with $\U(1)$ charge $(N-4)/(N-2)$ and (ii) singlets of $\SU(2)$.
Then, as in sec.~\ref{sec:2}, it is straightforward to see that the anti-symmetric representation contributes
$\SU(N-2)$ invariant states given as
\beq
\ket{ \pm (N-4)/2, 0, \pm (N-2)/2 }_a
\eeq
where $\pm (N-4)/2$ is the $\U(1)$ gauge charge, $0$ is the $\U(1)_B$ baryon charge which assings charge $1$ only to the hypermultiplets in the fundamental representation,
and $\pm (N-2)/2$ is the charge under the global symmetry $\U(1)_A$
which assigns charge $1$ only to the hypermultiplet in the anti-symmetric representation.

The possible gauge invariant states are, using the notation of sec.~\ref{sec:2}, given as
\beq
\ket{E_+} := &\ket{(ij)} \otimes \ket{(- N+\kappa) }_g \otimes \ket{\wedge^{+ N/2+2-\kappa+N_f/2}, ( + N/2+2-\kappa), ( + N/2+2-\kappa)}_h 
~\nonumber \\
& \otimes \ket{ + (N-4)/2, 0, + (N-2)/2 }_a~~~~~
 (\text{only if ~} |+ N/2+2-\kappa| \leq N_f/2) .  \label{eq:antip}  \\
\ket{E_-} := &\ket{(ij)} \otimes \ket{(+ N+\kappa) }_g \otimes \ket{\wedge^{- N/2-2-\kappa+N_f/2}, ( - N/2-2-\kappa), ( - N/2-2-\kappa)}_h ~\nonumber \\
& \otimes \ket{ - (N-4)/2, 0, - (N-2)/2 }_a~~~~~
(\text{only if ~} |- N/2-2-\kappa| \leq N_f/2) .\label{eq:antim}
\eeq
and also for small $N$,
\beq
\ket{F_+} := &\ket{(ij)} \otimes \ket{(- N+\kappa) }_g \otimes \ket{\wedge^{+ 3N/2-2-\kappa+N_f/2}, ( + 3N/2-2-\kappa), ( + 3N/2-2-\kappa)}_h 
~\nonumber \\
& \otimes \ket{ - (N-4)/2, 0, - (N-2)/2 }_a~~~~~
 (\text{only if ~} |+ 3N/2-2-\kappa| \leq N_f/2) .  \label{eq:antip}  \\
\ket{F_-} := &\ket{(ij)} \otimes \ket{(+ N+\kappa) }_g \otimes \ket{\wedge^{- 3N/2+2-\kappa+N_f/2}, ( - 3N/2+2-\kappa), ( - 3N/2 + 2-\kappa)}_h ~\nonumber \\
& \otimes \ket{ + (N-4)/2, 0, - (N-2)/2 }_a~~~~~
(\text{only if ~} |- 3N/2+2-\kappa| \leq N_f/2) .\label{eq:antim}
\eeq

Denoting the $\U(1)_A$ charge as $A$, the charges of $\ket{E_\pm} $ under $(I, B+\kappa I, A)$
are given as $(1, \pm (N +4)/2, \pm (N-2)/2)$, and those of $\ket{F_\pm} $ are given as $(1, \pm 3(N +4)/2, \mp (N-2)/2)$.
Note that if $N$ is large enough, the states $\ket{F_\pm} $ cannot exist because of the constraint $|\pm (3N/2-2)-\kappa| \leq N_f/2$.

In the present case, there are several $\U(1)$ symmetries, and hence it is more nontrivial to determine the Cartan generators of enhanced $\SU(2)_\pm$.
For this purpose, we use the argument of sec.~\ref{sec:pre} which uses RG flows between different theories.
We concentrate our attention to $\ket{E_\pm}$, since $\ket{F_\pm}$ do not generically exist.

First, we take $N+4-2\kappa=N_f$, in which case $\ket{E_+}$ is a singlet of $\SU(N_f)$ and hence we get enhanced $\SU(2)_+$
symmetry. Second, we give mass to the hypermultiplet in the anti-symmetric representation and integrate it out.
We assume that $\SU(2)_+$ symmetry is preserved under this RG flow if the sign of the mass is taken appropriately.
By integrating out the anti-symmetric matter, the Chern-Simons level is shifted as $\kappa \to \kappa'= \kappa+ (N-4)/2$ if the sign
of the mass is taken appropriately. In the low energy, we get a theory with $N_f$ fundamental flavors and
Chern-Simons level $\kappa'=\kappa+ (N-4)/2=N-N_f/2$. This is the correct Chern-Simons level for the enhancement of $\SU(2)_+$ in the low energy
theory. Therefore, in the low energy theory the Cartan of $\SU(2)_+$ is given by
\beq
H_+ = \frac{1}{2} I+ \frac{B + (\kappa+(N-4)/2) I}{2N} +c(A -\frac{1}{2}(N-2)  I),
\eeq
where we have taken into account the fact that all the states in the low energy theory have $A-\frac{1}{2}(N-2)I=0$ and 
hence we have ambiguity represented by the constant $c$. This is because all the low energy fields have $A=0$, and only instantons have charge
$A=\frac{1}{2}(N-2)I$ which can be shown in the same way as \eqref{eq:BCS}.
By doing the same consideration for $\SU(2)_-$, we get
\beq
H_\pm = \frac{1}{2} I \pm \frac{B + (\kappa \pm (N-4)/2 )I}{2N} \pm c (A  \mp \frac{1}{2}(N-2) I),
\eeq
where we have used the fact that the coefficient of $A$ must be $\pm c$ by parity and charge conjugation (spurious) symmetries.

We also use a different type of RG flows.
We can go to the subspace of the Higgs branch by giving vevs to two of the fundamental quarks so that the gauge group is broken as 
$\SU(N) \to \SU(N-1)$. We lose two of the fundamental flavors by the Higgs mechanism, but get one additional fundamental from the anti-symmetric matter.
Thus we get a theory in which $N$ and $N_f$ are reduced by one,
and the condition of the enhanced $\SU(2)_+$ is still preserved
if this symmetry exists before Higgsing. In terms of gauge invariant operators, this RG flow is triggered just by meson vevs, 
and hence we can assume that the $\SU(2)_+$ is unchanged under this RG flow.

The low energy baryon symmetry is a mixture of the high energy baryon symmetry and $\U(1) \subset \SU(N)$ charge 
such that the nonzero vevs of the quarks are neutral under the new baryon symmetry. 
The $\U(1) \subset \SU(N)$ charge is given as
\beq
\diag(-1, \frac{1}{N-1},\frac{1}{N-1}, \cdots, \frac{1}{N-1}).
\eeq
Then the charges under $(B,A)$ of the low energy fields are the following. The $N_f-2$ quarks have charge $(B,A)=(\frac{N}{N-1}, 0)$.
One quark which comes from the anti-symmetric matter has charge $(B,A)=(-\frac{N-2}{N-1},1)$. 
The anti-symmetric matter has charge $(B,A)=(\frac{2}{N-1},1)$.
Now, we require that the charges of the $N_f-2$ quarks and one quark are the same under $H_+$.
This is because $H_+$ should be insensitive to quark flavors.\footnote{If we consider a flow from $N=4$ to $N=3$,
the low energy $\SU(2)_+$ would be mixed with the flavor symmetry and the argument here need modification.}
This requirement gives $\frac{1}{N-1}=-\frac{N-2}{(N-1)N} +2c$, so we get $c=1/N$. Therefore we finally get the Cartan generators
\beq
H_\pm = \frac{1}{4} I \pm \frac{B + \kappa I +2A  }{2N} .
\eeq
This is our formula for the Cartan generators.
Note that gauge invariant chiral operators constructed from one anti-symmetric matters and two anti-fundamental quarks
have charge $H_\pm=0$ by using the above result, so these operators are singlets of $\SU(2)_\pm$.

Now we can see the difference from the case of theories without an anti-symmetric matter.
The charges of $\ket{E_+}$ and $\ket{E_-}$ under $(H_+, H_-)$ are given as $(1, - 1/2)$ and $(-1/2, 1)$, respectively.
This means the following. If both of $\ket{E_+}$ and $\ket{E_-}$ exist, they form a single $\SU(3)$ algebra instead of $\SU(2)_+ \times \SU(2)_-$.
In other words, the two nodes of $\SU(2)_+$ and $\SU(2)_-$ are connected.

Now let us see what happens in some cases. When $ N_f < N+4 $, nothing so special happens compared to the case without
anti-symmetric matters, because only one of $\SU(2)_+$ or $\SU(2)_-$ can be enhanced. 
It is just like that the anti-symmetric matter contributes effectively $N-4$ flavors of quarks.

When $\kappa=0$ and $N_f=N+4$, we get the connection of $\SU(2)_+$ and $\SU(2)_-$ into $\SU(3)$ discussed above.
The total non-abelian flavor symmetry is $\SU(3) \times \SU(N_f)$. When $\kappa=1/2$ and $N_f=N+5$,
the $\ket{E_+}$ is in the fundamental representation of the flavor group $\SU(N_f)$. This means that the Dynkin diagrams
of $\SU(3)$ and $\SU(N_f)$ are connected to form a single Dynkin diagram of $\SU(N_f+2)$.

When $\kappa=0$ and  $N_f=N+6$, we actually get a 6d theory. The $\ket{E_+}$ and $\ket{E_-}$ are
in the fundamental and anti-fundamental representations of the flavor group $\SU(N_f)$, so we get a Dynkin diagram like \eqref{eq:2n2}.
However, now the two black nodes of \eqref{eq:2n2} are also connected, and hence we actually get an affine $\SU(N_f+1)$ diagram.
Therefore, this is a 6d theory with $\SU(N_f+1)$ flavor symmetry.

Finally let us briefly comment on the case $N=4$. In this case, the $\U(1)_A$ symmetry of the anti-symmetric matter is enhanced to $\SU(2)_A$, since
the anti-symmetric representation is a real representation in this case.
One can see that $\ket{E_\pm}$ and $\ket{F_\pm}$ form doublets of $\SU(2)_A$. Thus the $\SU(2)_+ - \SU(2)_A- \SU(2)_-$ are actually combined
into a single $\SU(4)$. The rest of the story is similar to the above generic case.

\paragraph{Quivers.}
Now we consider the case that anti-symmetric matters are introduced into quiver theories.
Suppose that a node $i$ has an anti-symmetric matter. When this node satisfies $w_i+(N_i-4) < \sum_j A_{ij} N_k$,
nothing special happens. The anti-symmetric matter is just regarded as contributing effectively $N_i-4$ flavors as mentioned above.
Then the results are similar to those of sec.~\ref{sec:less}.

An interesting thing happens when $\kappa_i=0$ and $w_i+(N_i-4)= \sum_j A_{ij} N_k$. In this case,
the nodes in ${\rm Dyn}_+$ and ${\rm Dyn}_-$ corresponding to the node $i$ is connected. This is due to the above phenomenon of connecting $\SU(2)_+$
and $\SU(2)_-$ into $\SU(3)$. The results are thus similar to those of sec.~\ref{sec:more}.

\subsection{Symmetric representation}
Next let us consider a hypermultiplet in the symmetric representation when $N \geq 3$. We will be brief.

We consider an $\SU(N)$ gauge theory with a hypermultiplet in the symmetric representation and $N_f$ fundamental flavors.
The difference from the case of anti-symmetric representation is that we have an additional triplet of $\SU(2)$
in the decomposition $\SU(N) \supset \SU(2) \times \SU(N-2) \times \U(1)$. This triplet has $\U(1)$ charge $2$.
From this triplet, we get complex four zero modes which are in the four dimensional representation of the $\Sp(2) \cong \SO(5)$ rotation symmetry.
This can be quantized similar to the case of $\SU(2)$ gauginos. There are three $\Sp(2) $-invariant states,
\beq
\ket{ \pm 4, \pm 2}_s,~~~\ket{ 0,0}_s,
\eeq
where the first numbers in the kets, $\pm 4$ and $0$ are the gauge $\U(1)$ charges of these states, and the second numbers in the kets are
the $\U(1)_S$ charge which assigns charge $1$ to the symmetric matter and zero to others.
We need to take tensor products of these states with the states obtained above.

As long as the condition $2N- (N_f+N+4)-2|\kappa| \geq -2$ is satisfied, we cannot construct an $\U(1)$ invariant state
using the state $\ket{0}_s$. We have to use $\ket{ \pm 4}_s$. Then, the story is very similar to the case of the anti-symmetric matter,
if we consider the symmetric matter to be contributing effectively $N+4$ flavors instead of $N-4$.
The Cartan generators are given by
\beq
H_\pm = \frac{1}{4} I \pm \frac{B+\kappa I +2 S}{2N}   ,
\eeq
where $S$ is the $\U(1)_S$ charge.
Embedding into quivers is also the same.

It would also be interesting to consider theories with multiple matters in the symmetric and/or anti-symmetric representations.
For example, we can consider a theory with one symmetric representation and one anti-symmetric representation with $\kappa=0$ and $N_f=0$.
In this case the Cartan generators are $H_\pm=\pm (S+A)/N$. Now the instanton number $I$ disappears from these Cartan generators.
The nodes of $\SU(2)_+$ and $\SU(2)_-$ are connected by two lines, and hence instanton operators give an affine $A_1$ diagram.
Therefore this is a 6d theory.

\section*{Acknowledgments}
The author would very much like to thank Yuji Tachikawa for very helpful discussions and advice.
The work of KY is supported in part by DOE Grant No. DE-SC0009988.

\bibliographystyle{JHEP}
\bibliography{ref}

\end{document}